\documentclass[letterpaper,american,reprint, aps, prx, superscriptaddress]{revtex4-1}

\usepackage{textcomp}
\usepackage{amsmath}
\usepackage{amssymb}
\usepackage{graphicx}

\usepackage{hyperref}
\hypersetup{
    colorlinks=true,
    linkcolor=blue,
    citecolor=blue,      
    urlcolor=blue,
} 
\usepackage{babel}

\begin{document}
\title{Evolution of magnetic order in van-der-Waals antiferromagnet FePS$_3$
through insulator-metal transition}
\author{Matthew J. Coak}
\affiliation{Department of Physics, University of Warwick, Gibbet Hill Road, Coventry CV4 7AL, UK}
\affiliation{Cavendish Laboratory, Cambridge University, J.J. Thomson Ave, Cambridge CB3 0HE, UK}
\author{David M Jarvis}
\affiliation{Cavendish Laboratory, Cambridge University, J.J. Thomson Ave, Cambridge CB3 0HE, UK}
\author{Hayrullo Hamidov}
\affiliation{Cavendish Laboratory, Cambridge University, J.J. Thomson Ave, Cambridge CB3 0HE, UK}
\affiliation{Navoi State Mining Institute, 27 Galaba Avenue, Navoi, Uzbekistan}
\affiliation{National University of Science and Technology \textquotedblleft MISiS\textquotedblright , Leninsky Prospekt 4, Moscow 119049, Russia}
\author{Andrew R. Wildes}
\affiliation{Institut Laue-Langevin, CS 20156, 38042 Grenoble C\'edex 9, France}
\author{Joseph A. M. Paddison}
\affiliation{Materials Science and Technology Division, Oak Ridge National Laboratory, Oak Ridge, TN 37831, USA}
\affiliation{Cavendish Laboratory, Cambridge University, J.J. Thomson Ave, Cambridge CB3 0HE, UK}
\author{Cheng Liu}
\affiliation{Cavendish Laboratory, Cambridge University, J.J. Thomson Ave, Cambridge CB3 0HE, UK}
\author{Charles R.S. Haines}
\affiliation{Department of Earth Sciences, University of Cambridge, Downing Street, Cambridge CB2 3EQ, UK}
\affiliation{Cavendish Laboratory, Cambridge University, J.J. Thomson Ave, Cambridge CB3 0HE, UK}
\author{Ngoc T. Dang}
\affiliation{Institute of Research and Development, Duy Tan University, 550000 Da Nang, Viet Nam}
\affiliation{Faculty of Natural Sciences, Duy Tan University, 550000 Da Nang, Vietnam}
\author{Sergey E. Kichanov}
\affiliation{Frank Laboratory of Neutron Physics, Joint Institute for Nuclear Research, 141980 Dubna, Russia}
\author{Boris N. Savenko}
\affiliation{Frank Laboratory of Neutron Physics, Joint Institute for Nuclear Research, 141980 Dubna, Russia}
\author{Sungmin Lee}
\affiliation{Center for Correlated Electron Systems, Institute for Basic Science, Seoul 08826, Republic of Korea}
\affiliation{Department of Physics and Astronomy, Seoul National University, Seoul 08826, Republic of Korea}
\author{Marie Kratochv\'ilov\'a}
\affiliation{Center for Correlated Electron Systems, Institute for Basic Science, Seoul 08826, Republic of Korea}
\affiliation{Department of Physics and Astronomy, Seoul National University, Seoul 08826, Republic of Korea}
\affiliation{Faculty of Mathematics and Physics, Department of Condensed Matter Physics, Charles University, Prague, Czech Republic}
\author{Stefan Klotz}
\affiliation{Sorbonne Universit\'e, IMPMC, CNRS, UMR 7590, 4 Place Jussieu, 75252 Paris, France}
\author{Thomas Hansen}
\affiliation{Institut Laue-Langevin, CS 20156, 38042 Grenoble C\'edex 9, France}
\author{Denis P. Kozlenko}
\affiliation{Frank Laboratory of Neutron Physics, Joint Institute for Nuclear Research, 141980 Dubna, Russia}
\author{Je-Geun Park}
\affiliation{Center for Correlated Electron Systems, Institute for Basic Science, Seoul 08826, Republic of Korea}
\affiliation{Department of Physics and Astronomy, Seoul National University, Seoul 08826, Republic of Korea}
\affiliation{Center for Quantum Materials, Seoul National University, Seoul 08826, Republic of Korea}
\author{Siddharth S. Saxena}
\affiliation{Cavendish Laboratory, Cambridge University, J.J. Thomson Ave, Cambridge CB3 0HE, UK}
\affiliation{National University of Science and Technology \textquotedblleft MISiS\textquotedblright , Leninsky Prospekt 4, Moscow 119049, Russia}
\date{\today}
\begin{abstract}
Layered van-der-Waals 2D magnetic materials are of great interest in fundamental condensed-matter physics research, as well as for potential applications in spintronics and device physics. We present neutron powder diffraction data using new ultra-high-pressure techniques to measure the magnetic structure of Mott-insulating 2D honeycomb antiferromagnet FePS$_3$ at pressures up to 183~kbar and temperatures down to 80~K. These data are complemented by high-pressure magnetometry and reverse Monte Carlo modeling of the spin configurations. As pressure is applied, the previously-measured ambient-pressure magnetic order switches from an antiferromagnetic to a ferromagnetic interplanar interaction, and from 2D-like to 3D-like character. The overall antiferromagnetic structure within the $ab$ planes, ferromagnetic chains antiferromagnetically coupled, is preserved, but the magnetic propagation vector is altered from $(0\:1\:\frac{1}{2})$ to $(0\:1\:0)$, a halving of the magnetic unit cell size. At higher pressures, coincident with the second structural transition and the insulator-metal transition in this compound, we observe a suppression of this long-range-order and emergence of a form of magnetic short-range order which survives above room temperature. Reverse Monte Carlo fitting suggests this phase to be a short-ranged version of the original ambient pressure structure - with a return to antiferromagnetic interplanar correlations. The persistence of magnetism well into the HP-II metallic state is an observation in seeming contradiction with previous x-ray spectroscopy results which suggest a spin-crossover transition.
\end{abstract}
\maketitle

\section*{Introduction}

The study of low-dimensional magnetism has long been of great interest in fundamental condensed matter physics and has led to numerous applications. Layered magnetic van-der-Waals materials - compounds with planes of magnetic ions separated by wide van-der-Waals bonded gaps - provide a wide assortment of ideal systems for investigating 2D magnetism. A particularly powerful approach to investigating the physics in low-dimensional magnetic systems is to tune the interactions and dimensionality of the lattice and to explore how the magnetic and electronic behaviours of the material evolve. A large volume of recent work has focused on thickness control, to tune towards the `true 2D' limit of the atomic monolayer \citep{Park2016,Ajayan2016,Samarth2017,Zhou2016,Burch2018}, with many interesting results. A complementary, and critical, approach is to tune the magnetic interactions from 2D to 3D by applying hydrostatic pressure - in a van-der-Waals material, applying pressure will predominantly compress the weakly bonded interplanar separation and push the lattice towards a 3D character. This can be viewed as moving from the `quasi-2D' starting point in the opposite direction to a thickness dependence study, and addresses the same fundamental physics questions. Both approaches are needed for a full understanding of dimensionality in these fascinating systems to be accomplished. The use of pressure as a tuning parameter unlocks a cleaner and more controllable approach to map complete phase diagrams than methods such as chemical substitution or uniaxial strains.

\begin{figure*}
\centering{}\includegraphics[width=1.5\columnwidth]{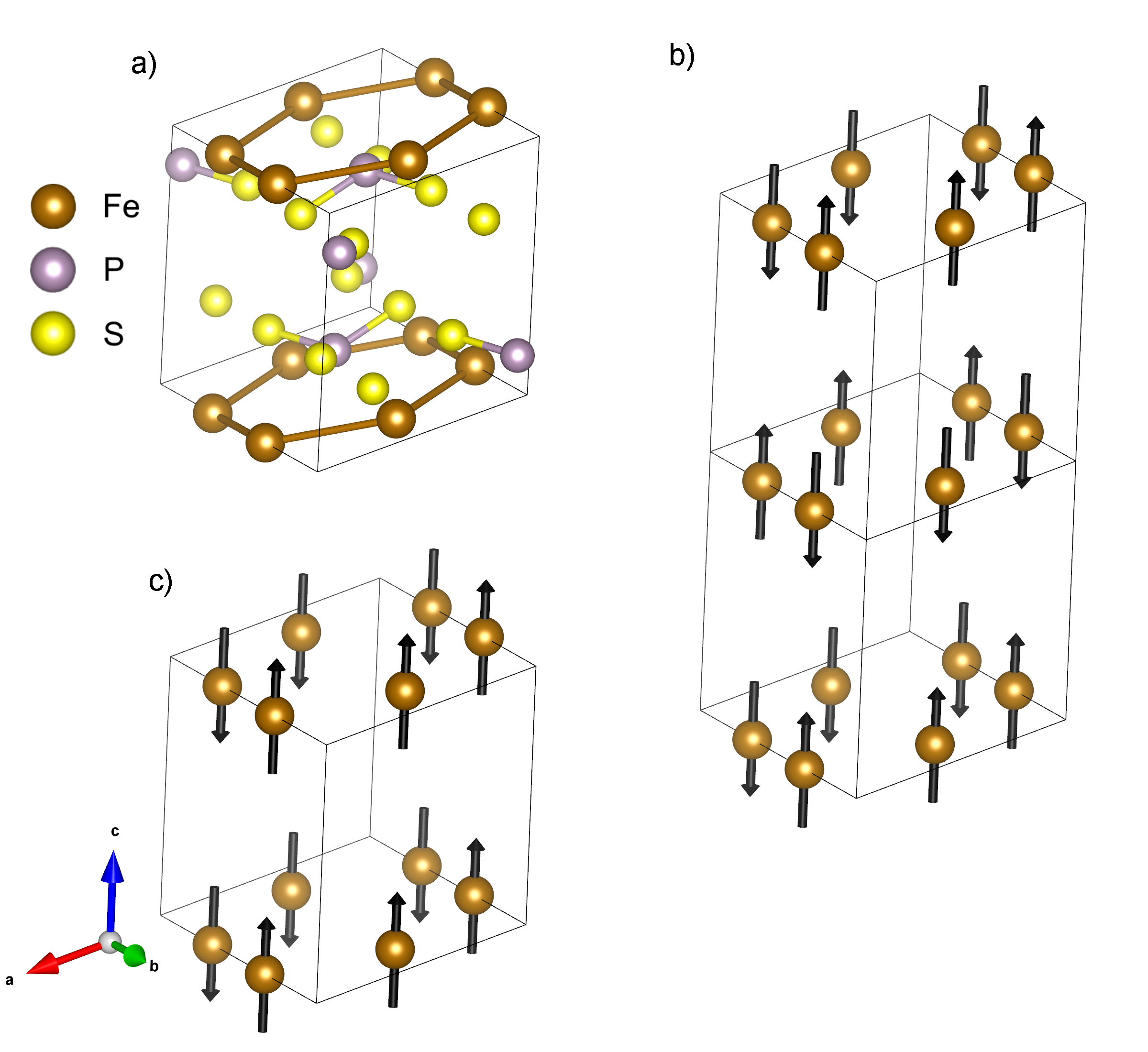}\caption{\label{fig:AmbientMagStructure}Crystal and magnetic structures of ambient pressure FePS$_3$. a) The monoclinic $C2/m$ crystal structure \citep{Ouvrard1985}. b) The magnetic cell of FePS$_3$ with the ambient-pressure $(0\:1\:\frac{1}{2})$ propagation vector - overall antiferromagnetic order comprised of zigzag chains along $a$, antiferromagnetically coupled to their neighbours and between the crystal planes \citep{Lancon_2016}. c) The magnetic unit cell with $(0\:1\:0)$ propagation vector, where planes are instead ferromagnetically aligned.}
\end{figure*}

The group of van-der-Waals (vdW) antiferromagnets $TM$PS$_3$ \citep{Grasso2002,Brec1986,Chittari2016,Rabu2003,Wang2018a}, with $TM$ a first-row transition metal, have been studied for some time as near-ideal 2D magnetic systems \citep{LeFlem1982,Kurosawa1983,Wildes1994,Wildes1998,Wildes1998a,Ronnow2000,Rule2002,Rule2003,Wildes2006,Rule2007a,Wildes2007,Rule2009,Wildes2012,Wildes2015,Lancon_2016,Wildes2017}. These compounds were first discovered in the 1890's and rigorously reported on in the 1980's \citep{Friedel1894,Friedel1894a,Klingen1968,Klingen1969,Klingen1970,Klingen1973,Ouvrard1985,Ouvrard1985a}. They all share the same basic crystal structure, shown in Fig \ref{fig:AmbientMagStructure} - a $C2/m$ monoclinic cell with a honeycomb arrangement of metal ions. These magnetic sites are arranged in slightly distorted hexagons in the $ab$ plane, separated by wide vdW gaps along the $c$ axis. Inter-site exchange both within the crystal planes and between layers
is mediated through P$_2$S$_6$ clusters surrounding the metal honeycombs, and there is significant competition between several exchange interactions of comparable strengths. Substituting the $TM^{2+}$ ion across the family leads to a wide variety of different magnetic order types and anisotropies, while keeping the same crystal structure, allowing clear and clean comparisons of, for example, Heisenberg versus Ising magnetic Hamiltonians in the 2D limit.

In recent years we have conducted a range of studies into the structural and transport properties of FePS$_3$, VPS$_3$, NiPS$_3$ and MnPS$_3$ under pressure \citep{Haines2018b,Coak2019b,Coak2019c,Jarvis2019,Coak2019d}. This work has revealed two structural transitions with increasing pressure common to all these materials, a shear motion of the $ab$ planes along the $a$ axis which brings the $\beta$ unit cell angle to close to 90\textdegree{} while remaining a monoclinic phase (structure HP-I), and then at higher pressures a collapse of the inter-planar separation and a transition to a trigonal symmetry group (HP-II) \citep{Haines2018b}. In FePS$_3$, the HP-I to HP-II structural transition is accompanied by an insulator-metal transition as electron overlap and hopping are greatly enhanced. This transition is a clear move from 2D towards a 3D character, as well as a change in the symmetry of the lattice, and can reasonably be expected to have a significant effect on the magnetism in this material. In the sister compound FePSe$_3$, continuing to increase pressure following the insulator-metal transition leads to the formation of a superconducting phase \citep{Wang2018} from the Mott-insulating 2D antiferromagnetic ambient state.

FePS$_3$ is an Ising antiferromagnet with a N\'eel temperature $T_{N}$ of around 118~K \citep{LeFlem1982}. Fig. \ref{fig:AmbientMagStructure}.b shows the established ambient-pressure magnetic structure \citep{Lancon_2016,Kurosawa1983} in the antiferromagnetic state. Zig-zag ferromagnetic chains with moments along $c^{*}$ are formed along $a$, ordered antiferromagnetically with their in-plane and interplane neighbours - magnetic propagation vector $(0\:1\:\frac{1}{2})$. However, the mere act of grinding a powder sample appears to introduce enough dislocations and stacking faults to essentially randomise long-range order along $c^{*}$, resulting in rod-like features in neutron diffraction patterns parallel to the $c^{*}$ axis and centred at the $(0\:1\:0)$ propagation vector \citep{Rule2003,Lancon_2016}. Single-crystal data do not display this behaviour, but rather a sharp, long-ranged 3D Bragg peak \citep{Lancon_2016}. A propagation vector of $(0\:1\:0)$ corresponds to much the same magnetic structure as the $(0\:1\:\frac{1}{2})$, but with the planes ferromagnetically coupled, as shown in Fig. \ref{fig:AmbientMagStructure}.c.

In this work we report the first high-pressure study of the magnetic structure in the $TM$PS$_3$ materials, through powder neutron diffraction. The development of double-toroidal sintered diamond anvils for the Paris-Edinburgh high-pressure press have allowed for measurements to be carried out above 100~kbar - we were able to observe the evolution of the magnetic structure well into the metallic HP-II phase of FePS$_3$ up to 183~kbar.

\section*{Methods}

Powder neutron diffraction patterns were collected on beamline D20 at the Institut Laue-Langevin, France \citep{Hansen2008}. Data were collected during two separate experiments. The first used a Paris-Edinburgh type pressure cell press with cubic boron nitride (cBN) anvils \citep{Klotz2005}, ILL DOI: \citep{ILL-DOI-D20Sept2018}. The second employed double-toroidal sintered diamond anvils in the same press \citep{Klotz2016}, to achieve significantly higher pressure values - ILL DOI: \citep{ILL-DOI-D20July2019}.
A single-crystal sample was ground into a fine powder in liquid nitrogen and in an argon glovebox to mitigate preferred orientation and water uptake. This powder was packed into the two halves of a Ti/Zr null scattering spherical gasket \citep{Marshall2002} and this placed between the anvils. The powder was wetted with 4:1 deuterated methanol/ethanol mixture to serve as a hydrostatic pressure medium. For the first experiment, a piece of lead was placed in the centre of the sample volume to act as a pressure gauge via its equation of state \citep{Straessle2014}. In the later experiment using sinterered diamond anvils, no pressure gauge was used, as the pressure dependence of the lattice parameters of the sample are known (at room temperature) \citep{Haines2018b} and so the pressure can be determined from the sample peak positions. The position of the $0\:0\:1$ peak was used for pressures up to 103~kbar, as this is highly sensitive to pressure, then the $1\:3\:1$ at pressures beyond this, as this then becomes more sensitive in the high-pressure structure. The pressures measured in the first experiment were 0, 20, 49, 72 and 80~kbar, and in the second 33, 78, 103, 157 and 183~kbar. Typical uncertainty on pressure values is a few kbar. All pressure changes were made above 300 K, then the cell cooled to 80~K and data taken at this base temperature and during warming, under constant load.

The wavelength of neutrons used was 2.42~\r{A}, via a graphite monochromator. Peaks in the neutron diffraction patterns originating from the anvils used, and the Pb manometer when present, were seen in addition to the sample signals. The diamond peaks are easily identified as they do not noticeably shift with pressure. With the diamond anvil measurements, a prominent peak at around 72 degrees of $2\theta$ was observed (removed from the presented datasets), but also a series of peaks at smaller $Q$. These did not move and hence must be attributed to the diamonds or other external cell environment, and could in fact be indexed as originating from diamond, but with a wavelength of $1/3$ that used in the measurement. The peaks are extremely weak compared to the principal diamond peak, so we conclude that a small weight of $\lambda/3$ neutrons were able to pass the graphite filter designed to suppress $\lambda/n$ wavelengths, and led to a measurable diamond diffraction pattern. These peaks are all marked as 'Diamond' in the presented data and can of course be ignored.

Additional neutron powder diffraction measurements at pressures up to 49~kbar were performed from 15 - 290~K on the DN-12 diffractometer, IBR-2 pulsed reactor, JINR, Russia \citep{Aksenov1999}. A powder sample of volume approximately 2~mm$^{3}$ was loaded into a sapphire anvil high pressure cell with culets of 4~mm diameter \citep{Somenkov2005}. Spherical cavities with a diameter of 2~mm were drilled at the culet centers to encourage quasi-hydrostatic pressure distribution at the sample surface. No pressure transmitting medium was used. The diffraction patterns were collected at scattering angles of 90\textdegree{} and 45\textdegree{} with resolution $\Delta d/d$ = 0.015 and 0.022, respectively. The pressure was measured using the ruby fluorescence technique \citep{Mao1986}. Pressure gradients were less than 10\% with respect to average pressure values. The diffraction patterns were analysed by the Rietveld method using the Fullprof software suite \citep{Rodriguez-Carvajal1993}.

The DC magnetisation data were measured in an MPMS SQUID magnetometer, Quantum Design, in an applied field of 0.1~T. The pressure cell used was a piston-cylinder clamp cell from Camcool Research Ltd, with a Daphne oil pressure medium and a lead manometer. Measurements were carried out to the maximum achievable pressure of cells of this type - around 10 kbar. Higher pressures would require a diamond anvil cell, where the tiny sample volume precludes resolving the small signal of an antiferromagnet like  FePS$_3$.

Reverse Monte Carlo fits were conducted in the Spinvert program \citep{Paddison2012,Paddison2013} to model short-range order in the highest pressure data. A supercell of 10x10x10 unit cells was used, and the previously solved HP-II crystal structure, with lattice parameters refined from the neutron data collected at this pressure.

\section*{Results}

\begin{figure*}
\centering{}\includegraphics[viewport=0bp 0bp 576bp 432bp,width=1.5\columnwidth]{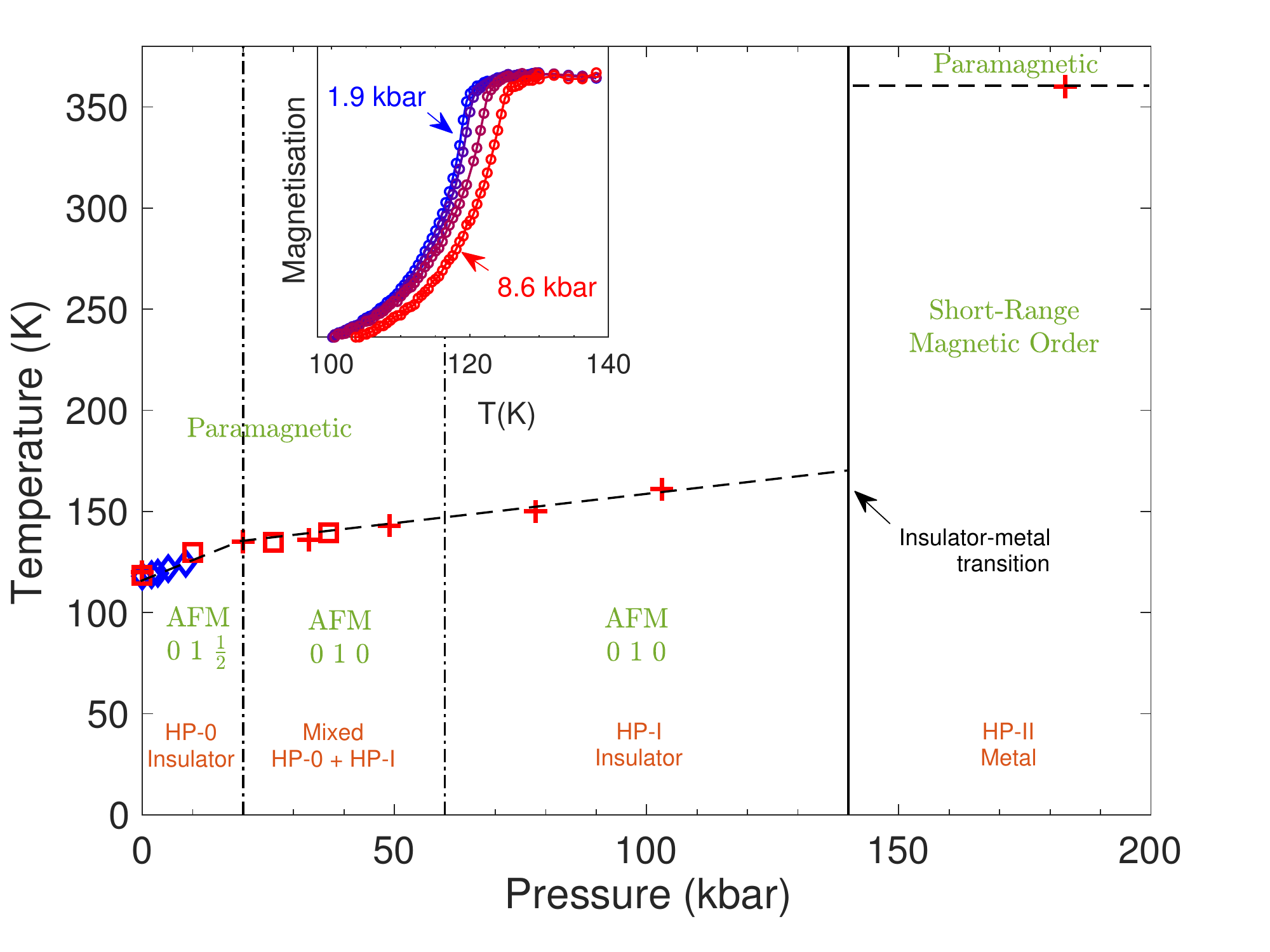}\caption{\label{fig:PhaseDiagram} Pressure-temperature phase diagram for FePS$_3$ constructed from data found in this work and the structural results from Ref \citep{Haines2018b}. Data points display extracted temperatures of magnetic transitions plotted against pressure. Blue circles show values extracted from magnetisation measurements, red crosses and squares from integrating the magnetic peaks and features in the neutron powder patterns from the ILL and JINR data respectively. Dashed lines delineating the AFM state are fits to the datapoints; gradients are stated in the text. Vertical lines on the phase diagram designate the pressures at which structural transitions occur. Inset shows normalised magnetisation data plotted against temperature for increasing pressures.}
\end{figure*}

\begin{figure*}
\centering{}\includegraphics[width=1.6\columnwidth]{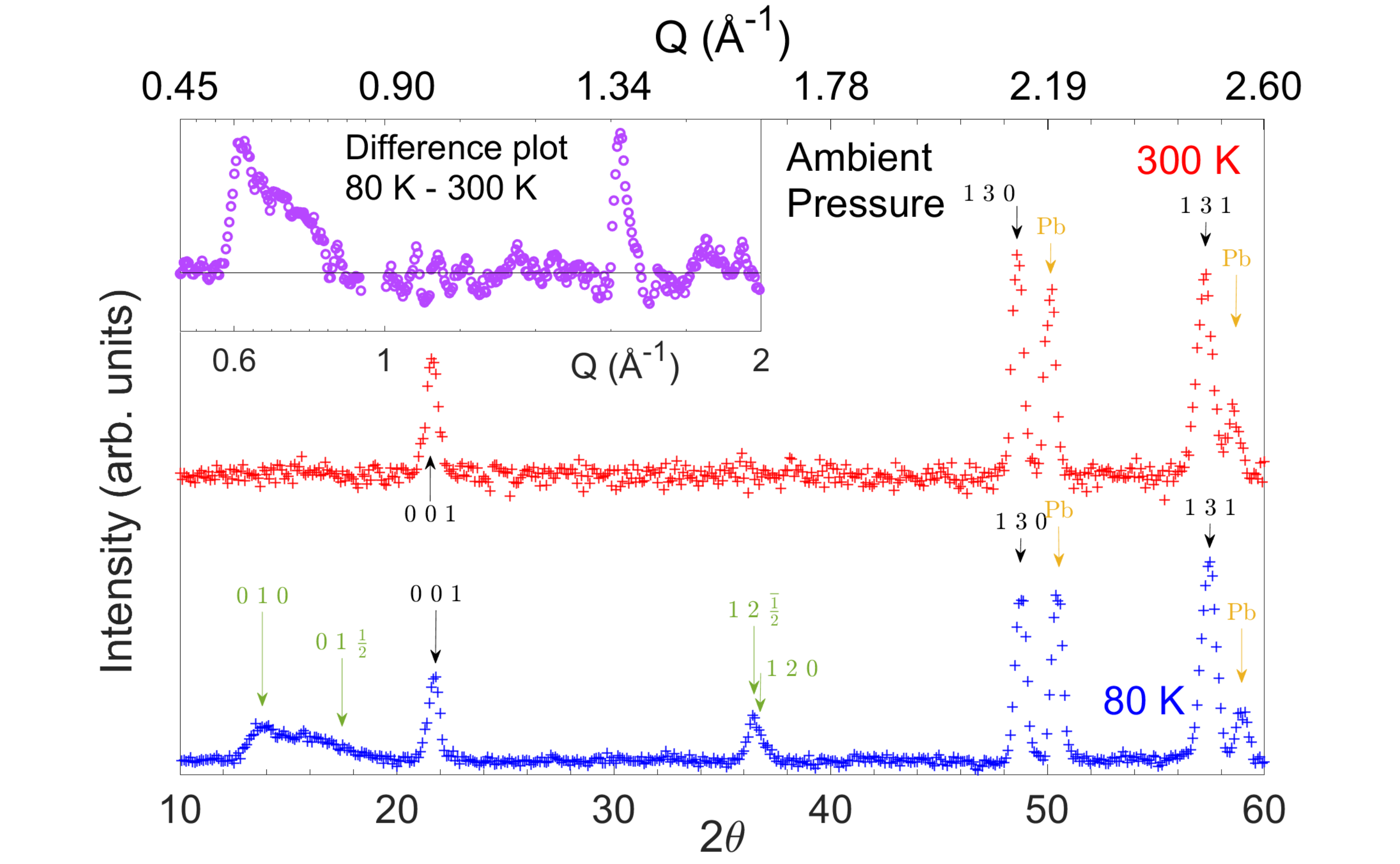}\caption{\label{fig:Ambient_HT-LT}Powder neutron diffraction patterns of FePS$_3$ at ambient pressure at 80~K and 300~K taken on D20, ILL. Nuclear and magnetic Bragg peaks are marked (magnetic in green) as indexed in the monoclinic LP phase. The magnetic propagation vector at ambient pressure is $(0\:1\:\frac{1}{2})$, but strain and stacking faults in the crystal due to grinding lead to 2D rod-like ordering with vector $(0\:1\:0)$, as described in Refs \citep{Kurosawa1983,Rule2003,Lancon_2016}. This causes the broad and asymmetric peak visible at small $Q$ in the 80~K data. An additional magnetic feature at higher $Q$ can be indexed as the $1\:2\:0$ and/or the $1\:2\:\bar{\frac{1}{2}}$. The broad diffuse background of the methanol-ethanol pressure medium has been subtracted from all plots shown. Inset shows the result of subtracting the 300~K data from the 80~K plotted against $Q$ to reveal the solely magnetic signal.}
\end{figure*}

The initial effect of pressure on the antiferromagnetic order in FePS$_3$ is to raise the value of $T_{N}$, as shown in Fig. \ref{fig:PhaseDiagram}. The inset shows magnetisation curves measured under pressure around the antiferromagnetic transition - these are clearly and continuously shifted upwards in temperature as pressure is increased. $T_{N}$ was extracted from differentiating these curves, and additionally from integrated magnetic peak intensities of the neutron diffraction patterns - these values are used to build the temperature-pressure phase diagram shown in the main figure. The value of $T_{N}$ increases linearly with pressure, initially at a rate of 0.77~K$.$kbar$^{-1}$. Pressure pushes the crystal planes together and hence will allow larger interlayer exchange, so it is consistent to see a strengthening of the magnetic order and it stabilising at higher temperatures. At approximately 20~kbar, there appears to be a change of slope, and thereafter the rate is reduced to 0.32~K$.$kbar$^{-1}$. This pressure corresponds to the value at which (at room temperature) the ambient pressure structure begins to (gradually) cross over to the HP-I structure \citep{Haines2018b}. A stiffening of the rate of change as the crystal enters the HP-I phase, which has stronger interplanar coordination, is again consistent and is reminiscent of the stiffening of vibrational modes seen in V$_{0.9}$PS$_3$ \citep{Coak2019c}. At higher pressures above 100~kbar, as will be discussed in the later sections, the magnetic order takes on completely different forms and the linear trend is no longer a valid description.

\begin{figure*}
\centering{}\includegraphics[width=1.6\columnwidth]{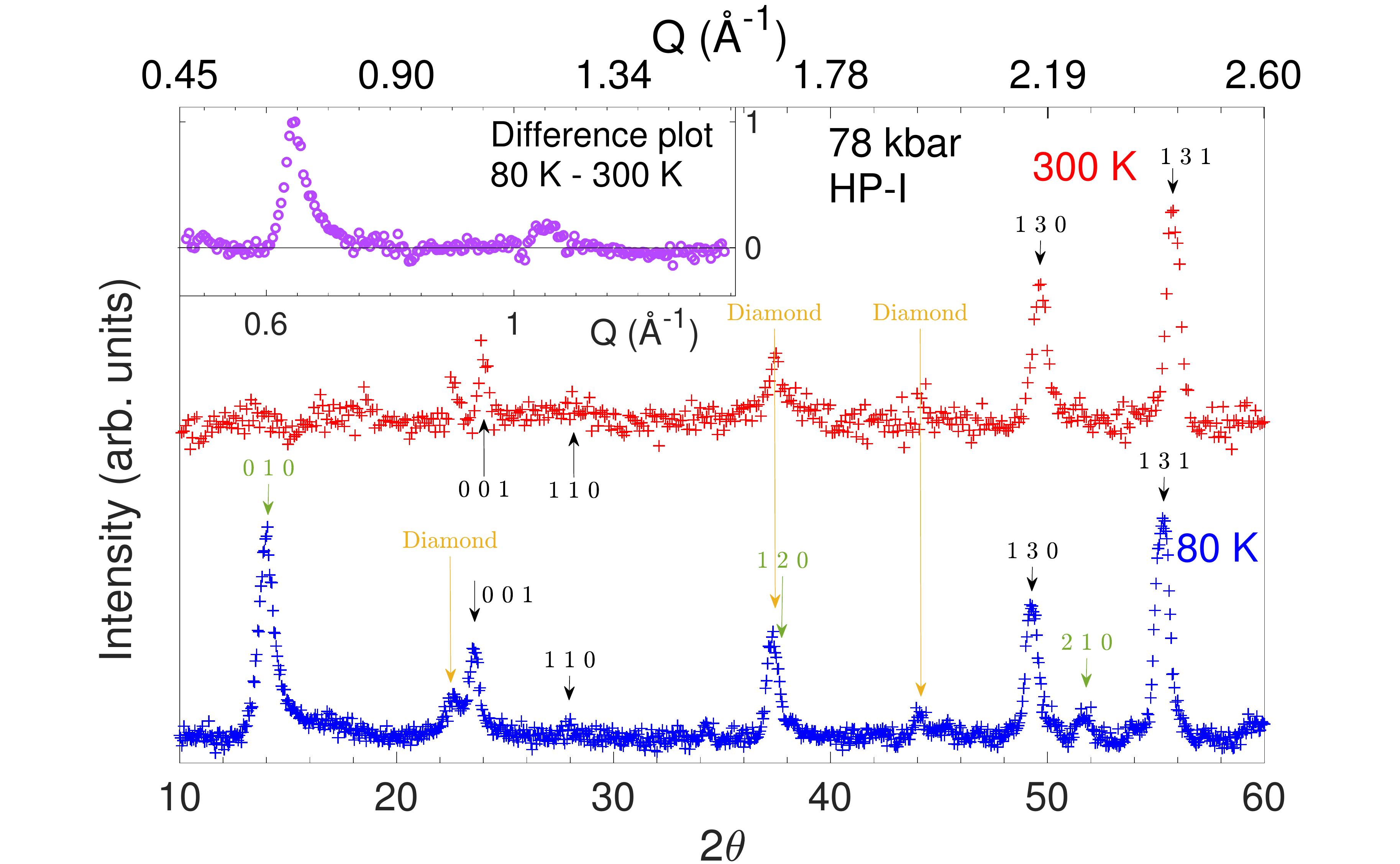}\caption{\label{fig:78kbar_HT-LT}Powder neutron diffraction patterns of FePS$_3$
at 78~kbar, at 80~K and 300~K taken on D20, ILL. Nuclear and magnetic Bragg peaks are marked as indexed in the monoclinic HP-I phase, and peaks due to the diamond anvils are marked in orange. The magnetic propagation vector is $(0\:1\:0)$ and characterised by a sharp symmetric Bragg peak. Additional magnetic peaks are seen at $1\:2\:0$ ( $(0\:1\:0)$ + $(1\:1\:0)$) and $2\:1\:0$ ( $(0\:1\:0)$ + $(2\:0\:0)$) from the combination of magnetic and nuclear scattering.  Inset shows the result of subtracting the 300~K data from the 80~K plotted against $Q$.}
\end{figure*}

\begin{figure}
\centering{}\includegraphics[width=0.9\columnwidth]{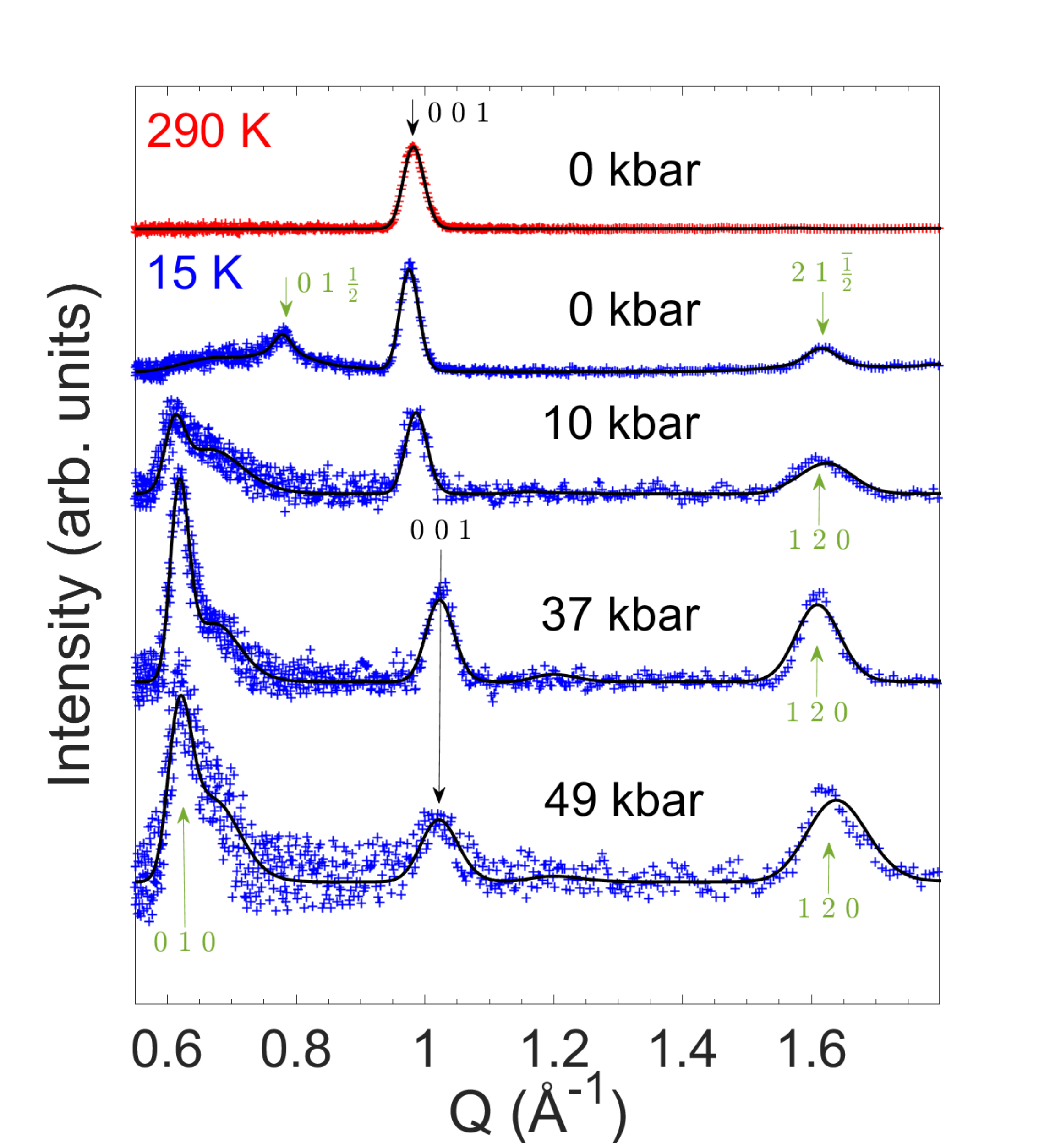}\caption{\label{fig:RussianData}Neutron diffraction patterns of FePS$_3$, measured at selected pressures up to 49~kbar at room temperature and 15~K. These data were taken in sapphire anvil cells at JINR, Russia. Crosses show the experimental data and black solid lines the results of refinements of the data by the Rietveld method. In the ambient pressure data shown at the top, the magnetic propagation vector is much more clearly $(0\:1\:\frac{1}{2})$. This is rapidly suppressed in favour of $(0\:1\:0)$, with an additional shoulder to the right of this peak suggestive of remnants of 2D-like ordering. Between 10 and 37~kbar, the $0\:0\:1$ nuclear Bragg peak clearly shifts due to the ambient - HP-I structural transition.}
\end{figure}

The ambient-pressure neutron powder diffraction patterns, obtained within the cBN single toroidal cell, are shown in Fig. \ref{fig:Ambient_HT-LT} for temperatures of 300~K and 80~K (the lowest temperature measured). The patterns have been normalised and a diffuse background due predominantly to the methanol-ethanol pressure medium has been subtracted from each data set by fitting a cubic spline function to the data between peaks and features. Below $T_{N}$ the magnetic peaks and features indexed and labeled in green in the figure are seen to appear. These match the form reported in previous work \citep{Rule2003,Lancon_2016}. A broad asymmetric feature, indicative of order of a 2D nature, spanning $0\:1\:0$ to $0\:1\:\frac{1}{2}$ is the principal magnetic feature. An additional peak at higher $Q$ can be indexed either as the $1\:2\:0$, corresponding to a combination of the magnetic $(0\:1\:0)$ with the nuclear $(1\:1\:0)$ peaks, or as the $1\:2\:\bar{\frac{1}{2}}$, a combination of $(0\:1\:\frac{1}{2})$ and $(1\:1\:\bar{1})$ - these have extremely similar $d$-spacings.

When pressure is applied, the magnetic structure shows a clear change, as illustrated in the 78~kbar example shown in Fig. \ref{fig:78kbar_HT-LT}. At this pressure FePS$_3$ will be in its HP-I structural phase, where the crystal planes have sheared and $\beta$ is close to 90 degrees. A single sharp magnetic Bragg peak is observed at $0\:1\:0$, replacing the rod-like scattering observed at ambient pressure. A move to a clear long-range magnetic order is consistent with the expected driving towards a 3D nature of the crystal. A $(0\:1\:0)$ magnetic propagation vector means that, as shown in Fig. \ref{fig:AmbientMagStructure}, the magnetic unit cell is now halved in volume, and the crystal planes coupled ferro-, rather than antiferro-magnetically. With the shear motion in the ambient - HP-I structural transition it is reasonable to expect changes to the inter-layer exchange pathways, consistent with this result. Only a minor change in exchange pathways is required to make this change, as the interlayer exchange under ambient pressure is very small (and negative) \citep{Lancon_2016} so could easily be driven to be small and positive. This is additionally consistent with the appearance of the rod-like features centred at $0\:1\:0$ in the ambient pressure patterns upon grinding - if the interlayer exchange is sensitive to shear along the $a$ axis and can easily be flipped in sign, grinding the sample would drive it to adopt a $(0\:1\:0)$ propagation vector. Though the literature \citep{Rule2003,Lancon_2016} had previously suggested stresses due to grinding to be responsible for the breaking of long-ranged order along $c^*$ and the reversal of some interplanar correlations, it is only the combination of our studies of how the structure and magnetic diffraction patterns evolve with pressure that allow us to now evidence this effect, and link these ideas together. The interplay of strain, structure and magnetic order shown here is potentially of great significance in the rapidly-growing field of heterostructure and device physics employing these materials, where strains are either an inescapable issue or a powerful engineering opportunity.

Data taken in sapphire anvil cells at JINR, Russia show more detail of the described low-pressure evolution of the magnetic structure - Fig. \ref{fig:RussianData}. In these data, the ambient pressure magnetic order is more clearly characterised by the $(0\:1\:\frac{1}{2})$ propagation vector than in the data shown in Fig. \ref{fig:Ambient_HT-LT}, with an almost symmetric 3D-like Bragg peak at this position. This can be explained by less grinding and shear strain during preparation of the sample than in the data taken at ILL. Reitveld refinements of these data yield a magnetic moment at ambient pressure of 5.40(5) $\mu_B$ per Fe, in line with previously published results \citep{Kurosawa1983}. Between the 10 and 37~kbar datasets shown, the ambient to HP-I structural transition occurs, shifting the position of the $0\:0\:1$ nuclear peak. Before this occurs, by 10~kbar, the magnetic order at $0\:1\:\frac{1}{2}$ has been suppressed in favour of a growing peak at the $0\:1\:0$ position, which reaches its full strength by 37~kbar. This is consistent with the earlier data shown, and this change has been discussed. Clearly visible in these data however is a shoulder or broad bump to the right of the $0\:1\:0$ peak. We interpret this as an asymmetric peak shape of the $0\:1\:0$ due to 2D-like magnetic ordering, as previously discussed in the context of the ambient pressure data in Fig. \ref{fig:Ambient_HT-LT}. This suggests the crystal, while gaining stronger inter-planar interactions and moving towards three-dimensionality, still maintains a quasi-2D character at these pressures.

\begin{figure*}
\centering{}\includegraphics[width=1.8\columnwidth]{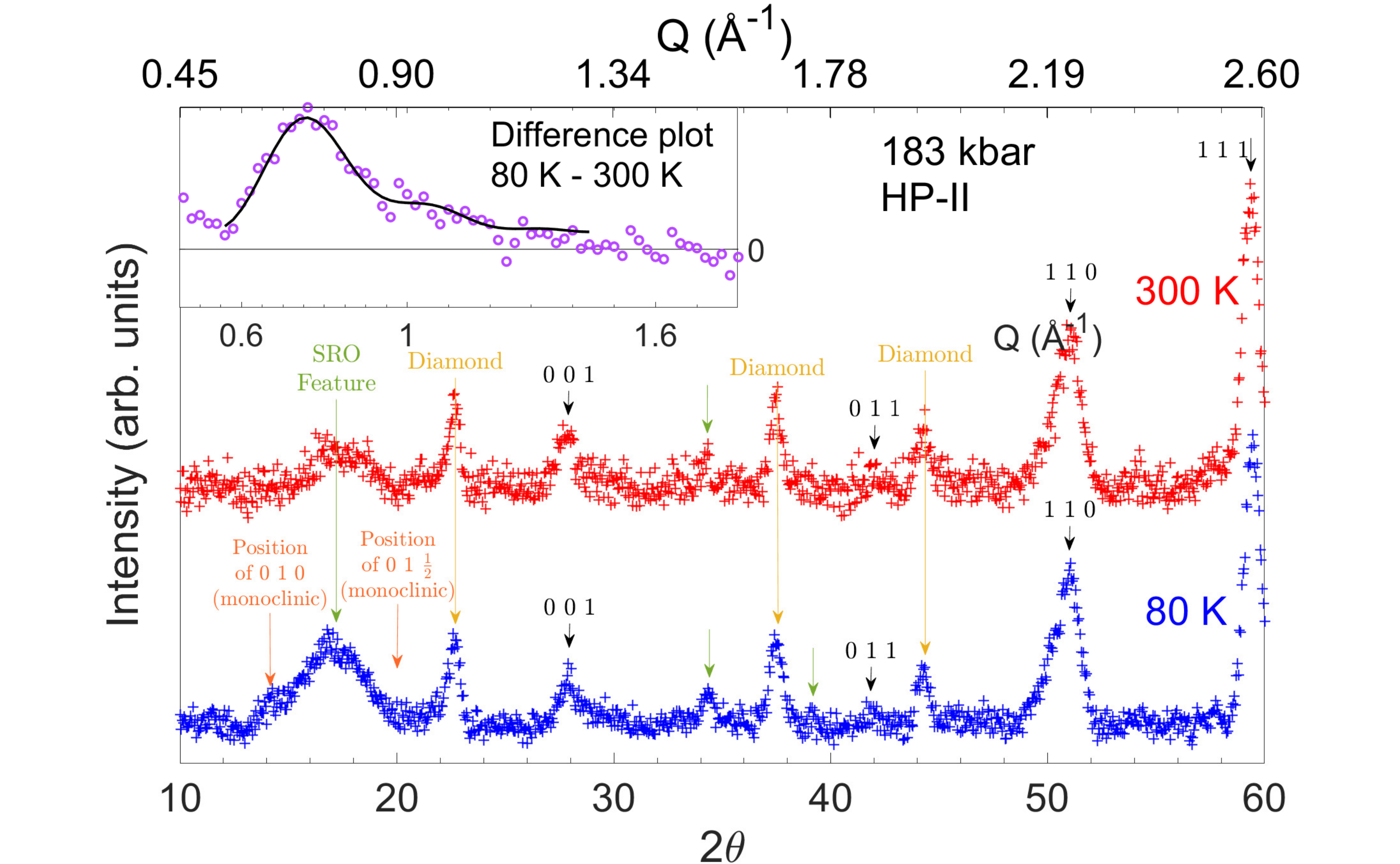}\caption{\label{fig:183kbar_HT-LT}Powder neutron diffraction patterns of FePS$_3$ at 183~kbar at 80~K and 300~K taken on D20, ILL. Peaks originating from the sintered diamond anvils are marked in orange. Nuclear Bragg peaks are marked as indexed in the trigonal HP-II phase. Broad temperature-dependent features indicative of short-range magnetic order are labeled in green. In dark orange are marked the positions the magnetic order peaks which would be described by $(0\:1\:0)$ and $(0\:1\:\frac{1}{2})$ in the monoclinic symmetry of the HP-I phase for ease of comparison to earlier figures. These same magnetic structures will of course have different indexing in the trigonal symmetry, but the $d$-spacings remain as shown. Inset shows the result of subtracting the 300~K data from the 80~K plotted against $Q$ - the SRO feature grows as temperature is decreased. The black solid line through these data shows the fit to this feature described in the text.}
\end{figure*}

At the highest pressures, within the HP-II trigonal structural phase and following metallisation, we observe another change in the magnetic ordering of FePS$_3$. Fig. \ref{fig:183kbar_HT-LT} shows the neutron diffraction patterns at room temperature and at 80~K for 183~kbar, the highest pressure measured. At this point the data show a broad hump or feature at low $Q$ typical of some kind of short-range magnetic order. This feature is reduced in magnitude upon increasing temperature, but is not suppressed completely upon reaching room temperature. Extrapolating the data taken while slowly warming would suggest extinction of this feature at 360~K, though due to the long count times required to gain reliable statistics at this point, there is a high degree of uncertainty on this prediction. Certainly whatever magnetic ordering this feature represents persists above room temperature. Additional peaks or features are also potentially seen at around 34.5\textdegree{} and 39\textdegree{}, but without sufficient quality of statistics in the data to be certain as to a temperature dependence. These cannot be indexed with any reasonable magnetic vector in the HP-II cell - but neither can the large hump feature at low $Q$.

\begin{figure}
\centering{}\includegraphics[width=0.9\columnwidth]{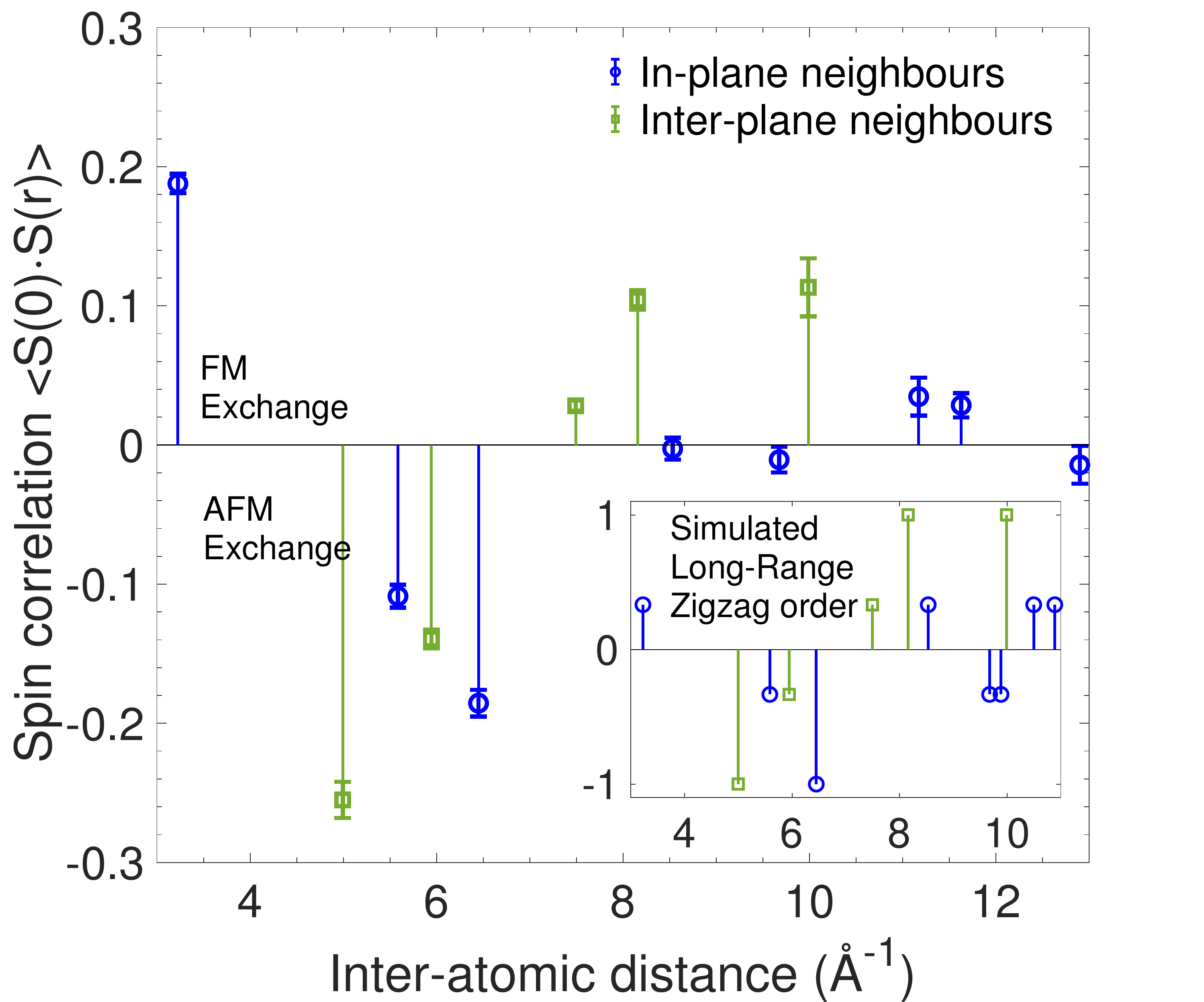}\caption{\label{fig:spinCorr}Normalised spin-spin correlation function $<\underline{S}(0)\cdot\underline{S}(r)>$ values for increasing inter-atomic distances from nearest-neighbour upwards for FePS$_3$ at 183~kbar. Perfect ferromagnetic correlations would give a value of +1, perfect antiferromagnetic -1. The values shown are extracted from the reverse Monte Carlo fits to the difference plot shown in the inset of Fig. \ref{fig:183kbar_HT-LT}. Nearest-neighbour correlations are seen to be ferromagnetic, and others predominantly antiferromagnetic, similar to those in ambient pressure FePS$_3$. The simulated correlation values for such ideal long-range zigzag order with antiferromagnetic interplanar correlations are shown in the inset for comparison.}
\end{figure}

In order to characterise the broad SRO feature seen at the highest pressures we carried out reverse Monte Carlo (RMC) refinements of spin configurations and fitted the results to the data. The calculations were carried out in the Spinvert program \citep{Paddison2012,Paddison2013}. The data used were the result of subtracting the room temperature data at 183~kbar from the 80~K data. The fits were restricted to the $Q$ region 0.56-1.50 \r{A}$^{-1}$ to capture only the SRO feature and avoid the Bragg peaks and the artefacts of the pressure cell background. The temperature-subtracted data and fit are shown in the inset of Fig. \ref{fig:183kbar_HT-LT} - good agreement is found over this range of $Q$. From this RMC fit, the spin correlation function could be extracted, $<\underline{S}(0)\cdot\underline{S}(r)>$, which quantifies the extent of ordering between a magnetic site and each of its neighbours - shown in Fig. \ref{fig:spinCorr}. These values are normalised such that perfect ferromagnetic correlations would give a value of 1, and perfect antiferromagnetic -1. A value of 0 implies no net correlation between the two sites, a complete lack of order over that length scale. We plot the first few unit cells' worth of correlation values extracted from the simulated spin arrangements; they die off to near-zero over the length shown as the order is short-ranged. As the figure shows, the spin correlations found are predominantly antiferromagnetic, except for ferromagnetic nearest-neighbour interactions. This appears remarkably similar to the ambient pressure $(0\:1\:\frac{1}{2})$ order (note that interplanar correlations are antiferromagnetic) - but a short-ranged and disordered version of that ferromagnetic-chain arrangement. The simulated correlation values for such ideal long-range zigzag order with antiferromagnetic interplanar correlations are shown in the inset of this figure for comparison. Note that this magnetic order would be indexed as $(0\:1\:\frac{1}{2})$ in the monoclinic structural phases, but is indexed as $(0\:\frac{1}{2}\:\frac{1}{2})$ in the trigonal HP-II structure found at this pressure. This means that the interplanar order and correlations have gone from antiferromagnetic in the ambient state, to ferromagnetic in the HP-I phase, and then back to antiferromagnetic in the HP-II crystal structure. As exchange is weak between the crystal planes, even small changes in the crystalline environment can flip the sign of the interplanar exhange integrals in this system.

Normalizing the data using the nuclear Bragg profile for this fitting process additionally yields a resulting effective magnetic moment of 6.6~$\mu_{B}$ per Fe ion, close to the ambient value of 5.7~$\mu_{B}$ \citep{Joy1992} within the sizable errors unavoidable in using this technique on high-pressure data. The uncertainties are complex and difficult to quantity exactly, as they are contributed to by imperfect background subtraction and absorption effects, but we estimate a ~20\% error on the moment value. Additionally, as the SRO is still present at 300~K, the temperature subtraction will have underestimated the magnitude of the feature - the true extracted moment we estimate to be even larger. We also note that additional details of the RMC results beyond those shown in Fig. \ref{fig:spinCorr} indicate that moment orientations have become far more isotropic at high pressure.

The observation of persisting magnetism, with a similar-magnitude moment, well into the HP-II phase is in disagreement with the x-ray emission spectroscopy results of Wang et al. \citep{Wang2018}, who report a spin crossover to an $S=0$ state at metallisation. The temperature dependence of the broad peak we observe in the high-pressure neutron data, and the fit results, preclude any origin other than magnetism for this feature - we do not observe a collapse of the Fe spin configuration, in apparent contradiction with Wang et al.'s findings, as well as some theoretical predictions, such as those of Zheng et al., \citep{Zheng2019}. However, as outlined in Refs \citep{Vanko2006,Rueff1999}, the x-ray spectra of the $K\beta$ Fe$^{\mathbf{II}}$peak are dependent not only on the ionic spin state, but also heavily on the local environment, geometry and surrounding bonds - including changes from insulating to metallic band structures. Further limits to the extraction of local moment from peak splitting as in this case are discussed in Ref \citep{Oh1992} - specifically it is only valid when the charge-transfer satellite is negligible. As the HP-I to HP-II transition entails not only an insulator-metal transition but also a significant shift in the crystal structure, local environment and dimensionality for the Fe ions, the change in the x-ray spectra seen by Wang et al. can potentially be explained by these effects, rather then a loss of spin and atomic magnetic moment - consistent with our observation of magnetism in the high-pressure phase.

We additionally note that an unchanged local moment on the Fe site would imply no valence or electron sharing changes on the Fe ions, and precludes any mixtures of high and low spin domains or fractional occupancy. It then seems likely that another ionic site is responsible for the metallic conduction - almost certainly the P sites. Phosphorus-phosphorus distances between the planes become significantly, abruptly, shorter in the HP-I to HP-II transition \citep{Haines2018b} to the point of reaching the typical length of a P-P bond. Formations of such bonds and hence valence changes and carrier donation from the P, rather than the Fe sites would seem a plausible explanation for a transition from insulating to metallic behaviour, while maintaining Fe local moments. This is in agreement with the findings in the recent theoretical treatment of the insulator-metal transition by Evarestov and Kuzmin \citep{Evarestov2020} which conclude that indeed phosphorus atoms are the main contributors to the conduction.

\begin{figure}
\begin{flushleft}a)\\\end{flushleft}
\includegraphics[width=0.95\columnwidth]{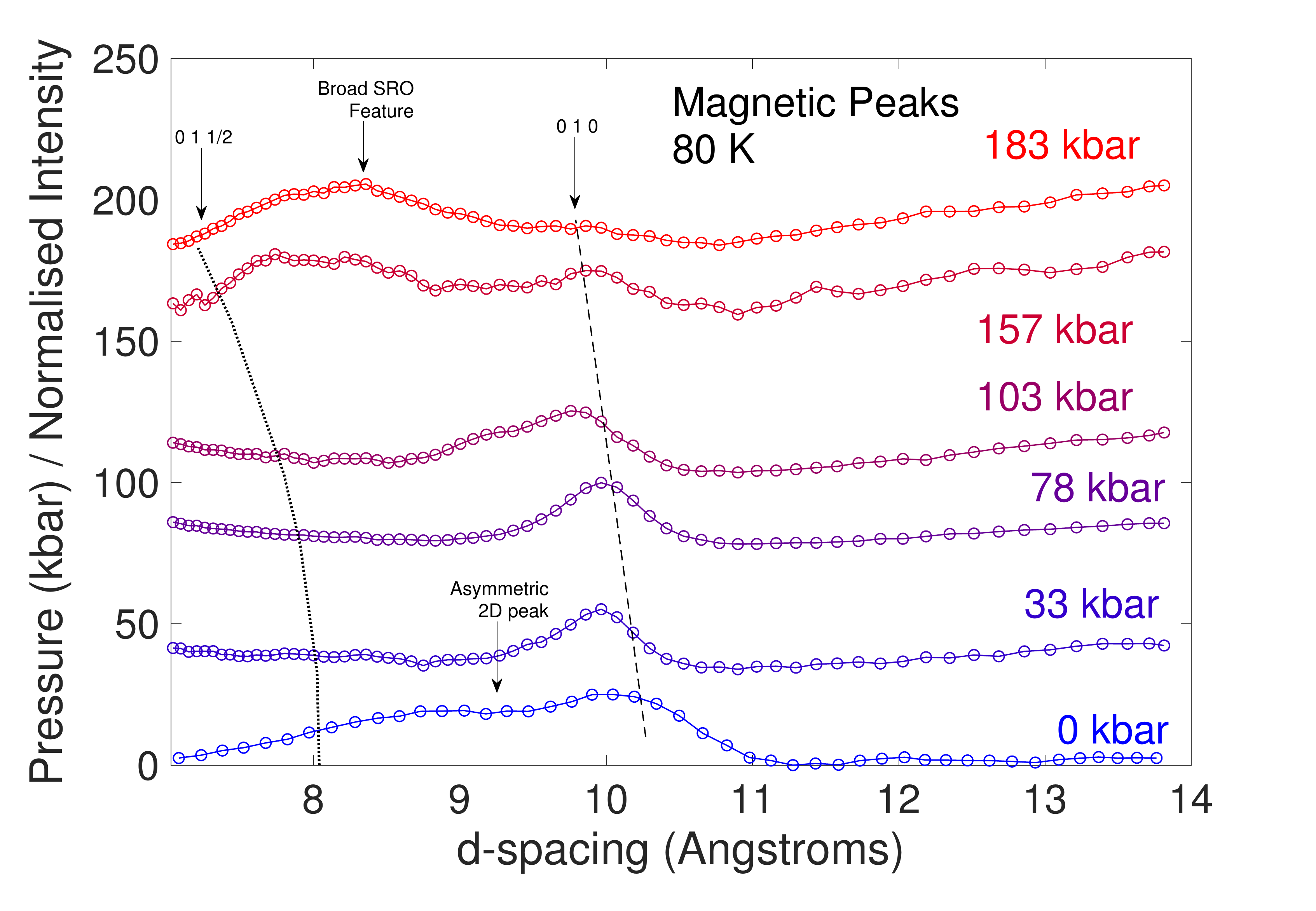}\\
\begin{flushleft}b)\\\end{flushleft}
\includegraphics[width=0.95\columnwidth]{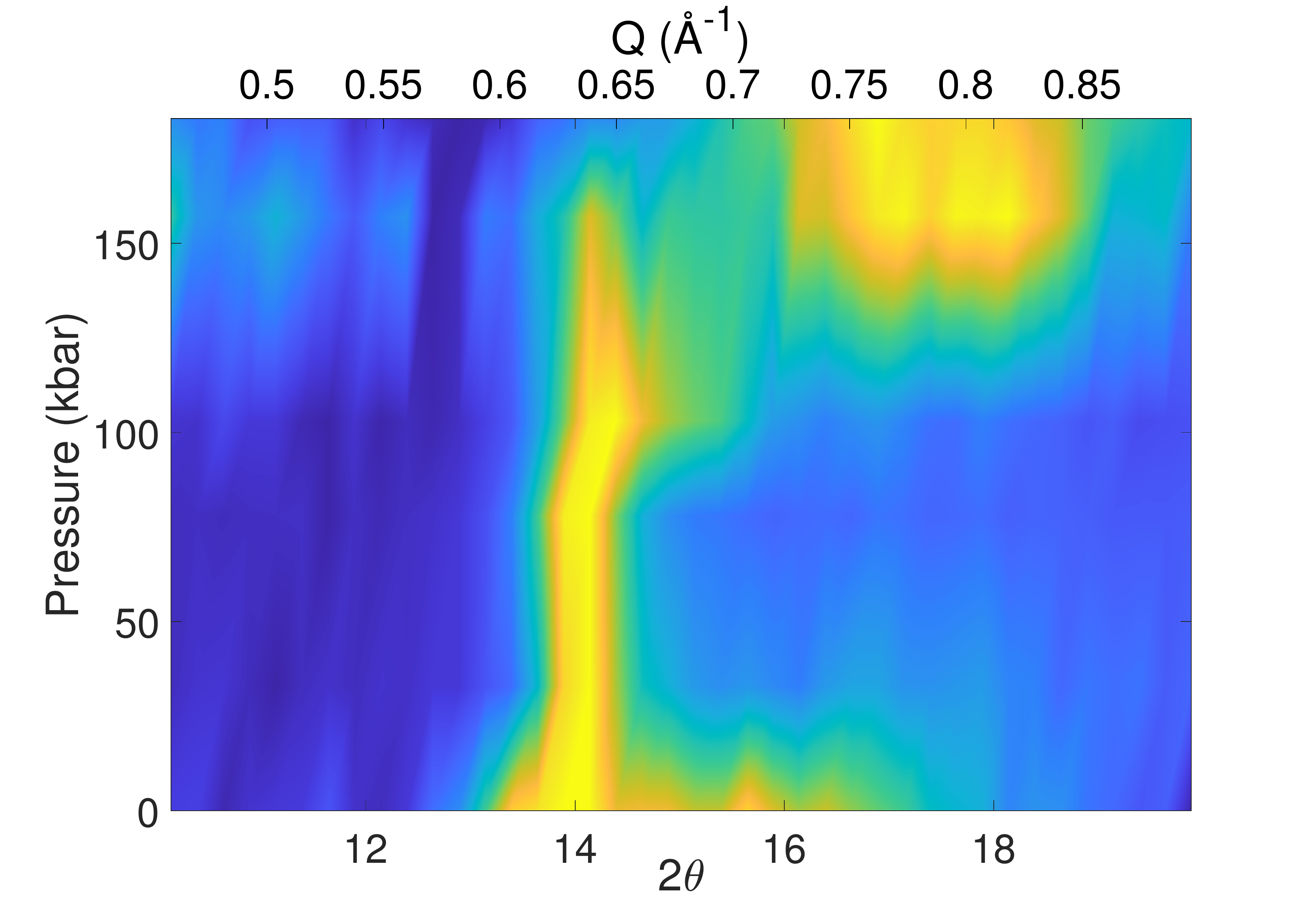}\caption{\label{fig:80K_P_Heatmap}
a) Magnetic peaks and features at 80~K for pressure points taken in sintered diamond cell on D20, ILL. The data have been normalised and their vertical intercept set to their corresponding pressure values. The pressure-dependent $d$-spacings of the $0\:1\:0$ and $0\:1\:\frac{1}{2}$ magnetic peak positions, extrapolated from the lattice parameters previously found for each pressure and indexed in the monoclinic space group, are shown as dashed and dotted lines. b) Heatmap of magnetic peaks at 80~K for sintered diamond pressures - colour represents normalised intensity. At ambient pressure the magnetic order is characterised by an asymmetric feature of both $0\:1\:0$ and $0\:1\:\frac{1}{2}$ contributions. This then sharpens into a sharp $0\:1\:0$ peak only by 20~kbar. The HP-I to HP-II structural transition can be seen around 150~kbar, which alters the magnetic state into a broad feature at higher $Q$ which persists above room temperature.}
\end{figure}

To summarise, Fig. \ref{fig:80K_P_Heatmap} displays the evolution of the principal magnetic features with pressure. Fig. \ref{fig:80K_P_Heatmap}a plots against $d$-spacing the diffraction patterns in the vicinity of the magnetic peaks, normalised and with their vertical axis offset set to their corresponding pressure values. Two changes are clear - at first the broad, asymmetric 2D-like peak spanning $0\:1\:0$ and $0\:1\:\frac{1}{2}$ sharpens into a single, sharper, 3D-like Bragg peak at the $0\:1\:0$ position. At higher pressures, this peak is suppressed, and a broad feature emerges at a lower $d$-spacing. The heatmap shown in Fig. \ref{fig:80K_P_Heatmap}b clearly shows these two changes in behaviour. The broad short-range-order feature at high $Q$ which emerges above around 130~kbar is seen to coexist for a time with the $0\:1\:0$ peak, one growing as the other is suppressed. This could, however, be due to pressure gradients and inhomogeneities within the sample region - a fraction of the sample may remain at an effective pressure below a transition the bulk has passed through. As the HP-I to HP-II structural transition is of strongly first-order character, if this change in magnetic order accompanies this transition one would expect it to be correspondingly abrupt.

\section*{Discussion}

We have measured the evolution of magnetic structure in Mott-insulating 2D honeycomb antiferromagnet FePS$_3$ up to 183~kbar using neutron powder diffraction and constructed the full pressure-temperature magnetic phase diagram, which corresponds well with previous structural studies. These results, the first of their kind, are likely to form the textbook example of dimensional tuning of magnetism in the crucial $TM$PS$_3$ systems and to lead to better understanding of the novel insulator-metal transitions and states in these compounds. Knowledge of the precise structure and the magnetic order is additionally essential for informing theoretical and computational studies of such materials. Beyond even this rich family of compounds, this work forms a fundamental study of the evolution of low-dimensional magnetism and the formation of exotic short-range ordered spin states in a near-ideal model system. 

Two changes in behaviour were observed. Initially, the ambient pressure magnetic order (antiferromagnetic interplanar coupling with strong 2D character) is switched to a ferromagnetic interplanar interaction, with fully 3D-like sharp Bragg peaks. The in-plane structure of ferromagnetic chains, antiferromagnetically coupled, is seemingly preserved - leading to overall antiferromagnetic order. The N\'eel temperature was found through high-pressure DC magnetometry as well as neutron diffraction to increase roughly linearly with pressure, by 36~K over 100~kbar, with a change in slope at the first structural transition. These observations are consistent with the increased interplanar exchange expected as the crystal planes are forced closer together and the structure pushed into the HP-I phase.

The second change to the magnetism is a gradual suppression of this $(0\:1\:0)$ order accompanied by an emergence of magnetic short-range order, with a transition or cutoff temperature seemingly above 300~K. This is indicated by a broad, roughly symmetric feature appearing at low $Q$ with strong temperature dependence. Reverse Monte Carlo fitting suggests this phase to be a short-ranged version of the original ambient pressure structure - with a return to antiferromagnetic interplanar correlations. This short-range order in the HP-II structure indicates frustration of the magnetism. This will result from competition in magnetic exchange and from the details of anisotropy on magnetic sites. The RMC fits suggest that the moments change direction considerably and become more isotropic, indicating that pressure and structural changes are having a significant effect on the crystal fields, and on the anisotropy.

The crossover to this short-range order state coincides with the previously observed HP-I to HP-II first-order structural transition and the corresponding insulator-metal transition in this material. The persistence of magnetism well into the HP-II metallic state is an observation in apparent contradiction with previous x-ray spectroscopy results which suggest a spin-crossover. We explain this discrepancy by suggesting the change in x-ray emission spectra to be due to the dramatic changes to the local environment of the Fe ion through the HP-I to HP-II transition and metallisation, rather than a loss of spin state. Additionally, a transition to metallic behaviour would imply a change in character of any magnetism to a more itinerant nature - it is not trivial to connect the observed change to SRO to this expectation. More microscopic details of this SRO state and its impact on properties such as electrical transport await future investigation - it is in fact possible that the high-pressure superconducting state in sister material FePSe$_3$ borders a similar magnetic phase.

\section*{Conclusions}

We have suceeded in mapping out the pressure-temperature magnetic phase diagram of van-der-Waals honeycomb antiferromagnet FePS$_3$: the first such study in the highly-scrutinised $TM$P$X$$_3$ materials. We find an intial flip and continuous strengthening in the interplanar exchange, which we discuss in terms of the previously observed shear structural transition in these materials at low pressures and a tuning from 2D to 3D character. The susceptibility of the material to this change in the magnetism - mechanical grinding of the powder is seemingly enough to instigate it to some degree - is crucial information for designers of few-layer devices and heterostructures seeking to incorporate the magnetic properties of $TM$P$X$$_3$, where strains and shear stresses are unavoidable. At higher pressures, where the material undergoes a collapse of the van-der-Waals gap and metallisation, we observe a loss of long-range magnetic order and the emergence of a short-range-ordered magnetic state in its place, which survives to much higher temperatures. This is counter to the previously-suggested collapse of the magnetic moment and hugely important information to aid in understanding both the nature of the exotic metallic state found in this region of the phase diagram and potentially the formation of unconventional superconductivity at higher pressures.

\begin{acknowledgments}
The authors would like to thank Suhan Son, Inho Hwang, Matt Tucker and Mary-Ellen Donnelly for their generous help and discussions. This work was supported by the Institute for Basic Science (IBS) in Korea (Grant No. IBS-R009-G1). We would also like to acknowledge support from Jesus College of the University of Cambridge and the CHT Uzbekistan programme. The work was carried out with financial support from the Ministry of Education and Science of the Russian Federation in the framework of Increase Competitiveness Program of NUST MISiS (K2-2017-024). J.-G.P. was partially supported by the Leading Researcher Program of the National Research Foundation of Korea (Grant No. 2020R1A3B2079375). J.A.M.P’s work at ORNL was supported by the U.S. Department of Energy, Office of Science, Basic Energy Sciences, Materials Sciences and Engineering Division (analysis of magnetic diffuse scattering data), and by ORNL LDRD 10004 (discussion of manuscript). This manuscript has been coauthored by UT-Battelle, LLC under Contract No. DE-AC05-00OR22725 with the U.S. Department of Energy. The United States Government retains and the publisher, by accepting the article for publication, acknowledges that the United States Government retains a non-exclusive, paid-up, irrevocable, world-wide license to publish or reproduce the published form of this manuscript, or allow others to do so, for United States Government purposes. The Department of Energy will provide public access to these results of federally sponsored research in accordance with the \href{http://energy.gov/downloads/doe-public-access-plan}{DOE Public Access Plan}. J.A.M.P.’s preliminary work at Cambridge was supported by Churchill College, University of Cambridge. This project has received funding from the European Research Council (ERC) under the European Union' s Horizon 2020 research and innovation programme (grant agreement No. 681260). Data presented in this paper resulting from the UK effort
will be made available at XXXXXXX.
\end{acknowledgments}

\bibliographystyle{apsrev4-1}

\begin{thebibliography}{57}

\bibitem{Park2016} 
  J.-G. Park, \href{\doibase 10.1088/0953-8984/28/30/301001}
  {
  J. Phys.: Condens. Matter \textbf{28}, 301001 (2016)
  }
  
\bibitem{Ajayan2016} 
  P. Ajayan and K. Banerjee, \href{\doibase 10.1063/pt.3.3297}
  {
  Physics Today \textbf{69}, 38 (2016)
  }
  
\bibitem{Samarth2017} 
  N. Samarth, \href{\doibase 10.1038/546216a}
  {
  Nature \textbf{546}, 216 (2017)
  }  
    
\bibitem{Zhou2016} 
  Y. Zhou, H. Lu and X. Zu and F. Gao, \href{\doibase 10.1038/srep19407}
  {
  Scientific Reports \textbf{6}, 19407 (2016)
  }
  
\bibitem{Burch2018} 
  K. S. Burch, D. Mandrus and J.-G. Park, \href{\doibase 10.1038/s41586-018-0631-z}
  {
  Nature \textbf{563}, 47 (2018)
  }    
  
\bibitem{Ouvrard1985} 
  G. Ouvrard, R. Brec and J. Rouxel, \href{\doibase 10.1016/0025-5408(85)90092-3}
  {
  Materials Research Bulletin \textbf{20}, 1181 (1985)
  }     
  
\bibitem{Lancon_2016} 
  D. Lan{\c{c}}on, H. C. Walker, E. Ressouche, B. Ouladdiaf, K. C. Rule, G. J. McIntyre, T. J. Hicks, H. M. R{\o}nnow and A. R. Wildes, \href{\doibase 10.1103/physrevb.94.214407}
  {
  Phys. Rev. B \textbf{94}, 214407 (2016)
  }   
  
\bibitem{Grasso2002} 
  V. Grasso and L. Silipigni, \href{http://adsabs.harvard.edu/abs/2002NCimR..25f...1G}
  {
  Rivista Del Nuovo Cimento \textbf{25}, 6 (2002)
  }   
  
\bibitem{Brec1986} 
  R. Brec, \href{\doibase 10.1016/0167-2738(86)90055-x}
  {
  Solid State Ionics \textbf{22}, 3 (1986)
  }    
  
\bibitem{Chittari2016} 
  B. L. Chittari, Y. Park, D. Lee, M. Han, A. H. MacDonald, E. Hwang and J. Jung, \href{\doibase 10.1103/PhysRevB.94.184428}
  {
Phys. Rev. B \textbf{94}, 184428 (2016)
  }      
  
\bibitem {Rabu2003}
P. Rabu and M. Drillon, \href {\doibase 10.1002/adem.200310082} 
{
Adv. Eng. Mater. \textbf{5}, 189 (2003)
}  
  
\bibitem {Wang2018a}
F. Wang, T. Shifa, P. Yu, P. He, Y. Liu, F. Wang, Z. Wang, X. Zhan, X. Lou, F. Xia and J. He, \href {\doibase 10.1002/adfm.201802151}
{
Advanced Functional Materials, \textbf{28}, 1802151 (2018)
}
  
\bibitem {LeFlem1982}  
G. Le~Flem,  R. Brec, G. Ouvard, A. Louisy and P. Segransan, \href{\doibase 10.1016/0022-3697(82)90156-1}
{
Journal of Physics and Chemistry of Solids \textbf {43}, 455 (1982)
}
  
\bibitem{Kurosawa1983} 
  K. Kurosawa, S. Saito, and Y. Yamaguchi, \href{http://jpsj.ipap.jp/link?JPSJ/52/3919/}
  {
  J. Phys. Soc. Jpn. \textbf{52}, 3919 (1983)
  }      
  
\bibitem{Wildes1994} 
  A. R. Wildes, S. J. Kennedy and T. J. Hicks, \href{\doibase 10.1088/0953-8984/6/24/002}
  {
  J. Phys.: Condens. Matter \textbf{6}, L335 (1994)
  }   
  
\bibitem {Wildes1998}
A.R. Wildes,  M. Harris and K. Godfrey, \href{\doibase 10.1016/s0304-8853(97)00666-5} 
{   
Journal of Magnetism and Magnetic Materials \textbf {177-181},  143 (1998)
}
  
\bibitem{Wildes1998a} 
  A. R. Wildes, B. Roessli, B. Lebech and K. W. Godfrey, \href{\doibase 10.1088/0953-8984/10/28/020}
  {
  J. Phys.: Condens. Matter \textbf{10}, 6417 (1998)
  }         
  
\bibitem {Ronnow2000}
H. R{\o}nnow,  A. R. Wildes and S. Bramwell \href{\doibase 10.1016/s0921-4526(99)01520-3} 
{   
Physica B: Condensed Matter \textbf{276-278},  676 (2000)
}
  
\bibitem {Rule2002}
K. Rule, S. Kennedy, D. Goossens, A. Mulders and T. Hicks \href {\doibase 10.1007/s003390201363} 
{   
Applied Physics A: Materials Science \& Processing \textbf {74},  s811 (2002)
}
  
\bibitem {Rule2003}
K. Rule, T. Ersez, S. Kennedy and T. Hicks, \href{\doibase 10.1016/s0921-4526(03)00179-0} 
{  
Physica B: Condensed Matter \textbf {335}, 6 (2003)
}
  
\bibitem {Wildes2006}
A. R. Wildes, H. M. R{\o}nnow, B. Roessli, M. J. Harris, and K. W. Godfrey \href {\doibase 10.1103/physrevb.74.094422} 
{  
 Phys. Rev. B \textbf {74}, 094422 (2006)
 }
  
\bibitem {Rule2007a}
 K. C. Rule, G. J. McIntyre, S. J. Kennedy and T. J. Hicks \href{\doibase 10.1103/physrevb.76.134402} 
 {
Phys. Rev. B \textbf {76}, 134402 (2007)
}
  
\bibitem{Wildes2007} 
  A. R. Wildes, H. M. R{\o}nnow, B. Roessli, M. J. Harris and K. W. Godfrey, \href{\doibase 10.1016/j.jmmm.2006.10.347}
  {
  Journal of Magnetism and Magnetic Materials \textbf{310}, 1221-1223 (2007)
  }    
  
\bibitem {Rule2009}
K. C. Rule, A. R. Wildes, R. Bewley, D. Visser, and T. J. Hicks \href {\doibase 10.1088/0953-8984/21/12/124214}
{
J. Phys.: Condens. Matter \textbf {21}, 124214 (2009)
}
  
\bibitem{Wildes2012} 
A. R. Wildes, K. C. Rule, R. I. Bewley, M. Enderle and T. J. Hicks, \href{\doibase 10.1088/0953-8984/24/41/416004}
  {
Journal of Magnetism and Magnetic Materials \textbf{310}, 1221-1223 (2012)
  }    
  
\bibitem{Wildes2015} 
A. R. Wildes, V. Simonet, E. Ressouche, G. J. McIntyre, M. Avdeev, E. Suard, S. A. J. Kimber, D. Lan{\c{c}}on, G. Pepe, B. Moubaraki, and T. J. Hicks, \href{\doibase 10.1103/physrevb.92.224408}
  {
Phys. Rev. B \textbf{92}, 224408 (2015)
  }  
  
\bibitem{Wildes2017} 
  A. R. Wildes, V. Simonet, E. Ressouche, R. Ballou and G. J. McIntyre, \href{\doibase 10.1088/1361-648X/aa8a43}
  {
  J. Phys.: Condens. Matter \textbf{29}, 455801 (2017)
  }  
  
\bibitem {Friedel1894}
M. Friedel, 
Bull. Soc. Chim. Fr. \textbf {11}, 115 (1894)
  
\bibitem {Friedel1894a}
M. Friedel,
C.R. Acad. Sci. \textbf {119}, 269 (1894)
  
\bibitem{Klingen1968} 
  W. Klingen, G. Eulenberger and H. Hahn, \href{\doibase 10.1007/bf00606219}
  {
  Die Naturwissenschaften \textbf{55}, 229 (1968)
  }  
  
\bibitem{Klingen1969}
W. Klingen,
Ph.D. thesis, {U}niversit\"at {H}ohenheim (LH) (1969)
  
\bibitem{Klingen1970} 
  W. Klingen, G. Eulenberger and H. Hahn, \href{\doibase 10.1007/BF00590690}
  {
  Die Naturwissenschaften \textbf{57}, 88 (1970)
  }   
  
\bibitem{Klingen1973} 
  W. Klingen, R. Ott and H. Hahn, \href{\doibase 10.1002/zaac.19733960305}
  {
  Zeitschrift f\"ur anorganische und allgemeine Chemie \textbf{396}, 271 (1973)
  }     
  
\bibitem{Ouvrard1985a} 
  G. Ouvrard, R. Fr\'eour, R. Brec and J. Rouxel, \href{\doibase 10.1016/0025-5408(85)90204-1}
  {
  Materials Research Bulletin \textbf{20}, 1053 (1985)
  }   
  
\bibitem{Haines2018b} 
  C. R. S. Haines, M. J. Coak, A. R. Wildes, G. I. Lampronti, C. Liu, P. Nahai-Williamson, H. Hamidov, D. Daisenberger and S. S. Saxena, \href{\doibase 10.1103/physrevlett.121.266801}
  {
  Phys. Rev. Lett. \textbf{121}, 266801 (2018)
  }  
  
\bibitem {Coak2019b}
M. J. Coak, S. Son, D. Daisenberger, H. Hamidov, C. R. S. Haines, P. L. Alireza, A. R. Wildes, C. Liu, S. S. Saxena and J.-G. Park, \href{\doibase 10.1038/s41535-019-0178-8}
{
npj Quantum Materials \textbf{4}, 38 (2019)
}
  
\bibitem {Coak2019c}
M. J. Coak, Y.-H. Kim, Y. S. Yi, S. Son, S. K. Lee and J.-G. Park, \href{\doibase 10.1103/physrevb.100.035120} 
{
Phys. Rev. B \textbf {100}, 035120 (2019)
}
  
\bibitem {Jarvis2019}
D. M. Jarvis, H. Hamidov, M. J. Coak, C. R. S. Haines, A. R. Wildes, C. Liu, D. Daisenberger and S. S. Saxena,
\emph{In preparation} (2020)
  
\bibitem {Coak2019d}  
M. J. Coak, D. M. Jarvis, H. Hamidov, C. R. S. Haines, P. L. Alireza, C. Liu, S. Son, I. Hwang, G. I. Lampronti, D. Daisenberger, P. Nahai-Williamson, A. R. Wildes, S. S. Saxena and J.-G. Park, \href {\doibase 10.1088/1361-648x/ab5be8} 
{
Journal of Physics: Condensed Matter \textbf {32}, 124003 (2019)
}
  
\bibitem{Wang2018} 
  Y. Wang, J. Ying, Z. Zhou, J. Sun, T. Wen, Y. Zhou, N. Li, Q. Zhang, F. Han, Y. Xiao, P. Chow, W. Yang, V. V. Struzhkin, Y. Zhao and H.-K Mao, \href{\doibase 10.1038/s41467-018-04326-1}
  {
  Nat. Comms \textbf{9}, 1914 (2018)
  }   
  
\bibitem{Hansen2008}
T. C. Hansen,  P. F. Henry, H. E. Fischer, J. Torregrossa and P. Convert, \href {\doibase 10.1088/0957-0233/19/3/034001} 
{
Measurement Science and Technology \textbf {19}, 034001 (2008)
}
  
\bibitem {Klotz2005}
S. Klotz, T.  Str\"asle, G.  Rousse, G.  Hamel and V.~ Pomjakushin, \href {\doibase 10.1063/1.1855419} 
{
Applied Physics Letters \textbf {86}, 031917 (2005)
}
  
\bibitem {ILL-DOI-D20Sept2018}
S. S. Saxena, M. J. Coak, S. E. Dutton, C. R. S. Haines, H.  Hamidov, T. Hansen, D. M. Jarvis, C.  Liu and A. R. Wildes, \href {\doibase 10.5291/ill-data.5-31-2545}
{
\emph{Effect of pressure on the magnetism of the two-dimensional antiferromagnet {FePS}$_3$}, ILL DOI (2018)
}

\bibitem {Klotz2016}
S. Klotz, T. Str\"asle, B. Lebert, M. d'Astuto and T. Hansen \href {\doibase 10.1080/08957959.2015.1136624}
 {
 High Pressure Research \textbf {36}, 73 (2016)
 }
  
\bibitem {ILL-DOI-D20July2019}
M. J. Coak, S. E. Dutton, C. R. S. Haines, H. Hamidov, T. Hansen, D. M. Jarvis, S. Klotz, C. Liu, J.-G. Park, S. S. Saxena and A. R. Wildes, \href {\doibase 10.5291/ill-data.5-31-2646} 
{
\emph{Effect of pressure on the magnetism of the two-dimensional antiferromagnet {FePS}$_3$}, ILL DOI (2019)
}
  
\bibitem {Marshall2002}
W. G. Marshall and D. J. Francis, \href {\doibase 10.1107/s0021889801018350} 
{
Journal of Applied Crystallography \textbf{35},  122 (2002)
}
  
\bibitem {Straessle2014}
T. Str\"assle, S. Klotz, K. Kunc, V. Pomjakushin and J. S. White, \href {\doibase 10.1103/physrevb.90.014101} 
{  
Phys. Rev. B \textbf {90}, 014101 (2014)
}
  
\bibitem {Aksenov1999}
V. Aksenov, A. Balagurov, V. Glazkov, D. Kozlenko, I. Naumov, B. Savenko, D. Sheptyakov, V. Somenkov, A. Bulkin, V. Kudryashev and V. Trounov, \href {\doibase 10.1016/s0921-4526(98)01392-1}
{
Physica B: Condensed Matter \textbf {265}, 258 (1999)
}
  
\bibitem {Somenkov2005}
V. A. Somenkov, \href {\doibase 10.1088/0953-8984/17/40/001} 
{
Journal of Physics: Condensed Matter \textbf {17}, S2991 (2005)
}
  
\bibitem {Mao1986}
H. Mao, J. Xu, and P. Bell, \href {\doibase 10.1029/jb091ib05p04673} 
{
Journal of Geophysical Research \textbf {91}, 4673 (1986)
}
  
\bibitem {Rodriguez-Carvajal1993}
J. Rodriguez-Carvajal,
Phys. B. \textbf {192}, 55 (1993)
  
\bibitem {Paddison2012}
J. A. M. Paddison and A. L. Goodwin, \href {\doibase 10.1103/physrevlett.108.017204}
{
Phys. Rev. Lett. \textbf {108}, 017204 (2012)
}
  
\bibitem {Paddison2013}
J. A. M. Paddison, J. R. Stewart and A. L. Goodwin, \href {\doibase 10.1088/0953-8984/25/45/454220}
{
Journal of Physics: Condensed Matter \textbf {25}, 454220 (2013)
}
  
\bibitem {Joy1992}
P. A. Joy and S. Vasudevan, \href {\doibase 10.1103/PhysRevB.46.5425}
{ 
Phys. Rev. B \textbf{46}, 5425 (1992)
}
    
\bibitem {Zheng2019}
Y. Zheng, X.-x. Jiang, X.-x. Xue, J. Dai and Y. Fengn, \href {\doibase 10.1103/physrevb.100.174102}
{ 
Phys. Rev. B \textbf{100}, 174102 (2019)
}

\bibitem {Vanko2006}
G. Vank\'o, T. Neisius, G. Moln\'ar, F. Renz, S. K\'arp\'ati, A. Shukla and F. M. F. de Groot, \href{\doibase 10.1021/jp0615961}
{
The Journal of Physical Chemistry B \textbf {110}, 11647 (2006)
}
  
\bibitem {Rueff1999}
J.-P. Rueff, C.-C. Kao, V. V. Struzhkin, J. Badro, J. Shu, R. J. Hemley and H. K. Mao, \href {\doibase 10.1103/physrevlett.82.3284}
{
Phys. Rev. Lett. \textbf{82}, 3284 (1999)
}
  
\bibitem {Oh1992}
S.-J. Oh, G.-H. Gweon and J.-G. Park, \href{\doibase 10.1103/physrevlett.68.2850} 
{
Phys. Rev. Lett. \textbf{68}, 2850 (1992)
}

\bibitem {Evarestov2020}
R.A. Evarestov and A. Kuzmin, \href{\doibase 10.1002/jcc.26178} 
{
Journal of Computational Chemistry. \textbf{41}, 14 (2020)
}

  
\end{thebibliography}

\end{document}